\documentclass[conference]{IEEEtran}
\IEEEoverridecommandlockouts
\usepackage{cite}
\usepackage{amsmath,amssymb,amsfonts}
\usepackage{algorithmic}
\usepackage{graphicx}
\usepackage{textcomp}
\usepackage{xcolor}
\usepackage{xcolor}
\usepackage{subscript}
\usepackage{float}
\usepackage{xspace}
\usepackage{multirow}
\newcommand{\accc}{$AC^3$\xspace}

\def\BibTeX{{\rm B\kern-.05em{\sc i\kern-.025em b}\kern-.08em
    T\kern-.1667em\lower.7ex\hbox{E}\kern-.125emX}}
\begin{document}

\title{
Harnessing Data Spaces to Build Intelligent Smart City Infrastructures Across the Cloud-Edge Continuum
\\
\thanks{This work is supported by the European Union’s Horizon
Program under the Agile and Cognitive Cloud edge Continuum management (AC3)
 project (Grant No. 101093129).}
}

\author{

\IEEEauthorblockN{Dimitrios Amaxilatis, Themistoklis Sarantakos,\\Nikolaos Tsironis}
\IEEEauthorblockA{\textit{Spark Works Ltd.}\\
Galway, Ireland \\
\{d.amaxilatis,tsarantakos,ntsironis\}@sparkworks.net\\
0000-0001-9938-6211,0000-0002-7517-6997,\\0009-0009-8084-613X}
\and

\IEEEauthorblockN{Souvik Sengupta\\~}
\IEEEauthorblockA{\textit{IONOS, SE}\\
Karlsruhe, Germany \\
souvik.sengupta@ionos.com\\
0000-0002-0056-1540}
\and 

\IEEEauthorblockN{Kostas Ramantas, Jhofre Ojeda\\~}
\IEEEauthorblockA{\textit{Iquadrat Informática, S.L.}\\
Barcelona, Spain \\
\{kramantas,j.ojeda\}@iquadrat.com\\
0000-0002-1304-784X}



}

\maketitle

\begin{abstract}
Smart cities are increasingly adopting data-centric architectures to enhance the efficiency, sustainability, and resilience of urban services.
This paper presents a real-world use case of intelligent infrastructure monitoring, developed and deployed within a data space–enabled cloud-edge framework.
Our use case demonstrates how edge computing, containerized microservices, and interoperable data sharing can be leveraged to address key challenges such as sensor integration, data privacy, network reliability, and system scalability.
By enabling localized analytics, real-time inference, and decentralized data exchange, the implementation supports critical urban applications including infrastructure condition assessment, environmental sensing, and public service optimization.
Data spaces serve as the backbone for secure, sovereign, and trusted collaboration among stakeholders, facilitating the controlled exchange and enrichment of sensor data.
The use case also showcases the training and deployment of AI/ML services directly at the edge, allowing for optimized resource use and timely decision-making.
Validated through a data space–integrated IoT testbed with advanced cloud–edge capabilities, this implementation highlights the transformative potential of combining AI, edge computing, and data spaces in future-ready smart city ecosystems.
\end{abstract}



\section{Introduction}
As urban environments grow increasingly complex, cities around the world are turning to digital technologies to enhance the efficiency, safety, and sustainability of public services.
At the heart of this digital transformation lies the smart city paradigm, where data-driven decision-making supports diverse functions such as traffic optimization, infrastructure maintenance, and environmental monitoring.

However, realizing the full potential of smart cities requires more than deploying connected devices—it demands trusted, interoperable systems that enable secure and sovereign data exchange among a wide range of stakeholders, including municipalities, infrastructure operators, private companies, and citizens.
Data spaces have emerged as a key enabler of this vision, providing federated, standardized environments that support data sovereignty, interoperability, and compliance with privacy regulations.
Complementing this, the cloud–edge continuum plays a critical role in realizing smart city deployments at scale. 
The vast number of connected devices and sensors, combined with the real-time and mission-critical nature of many urban services—such as traffic control, energy distribution, and emergency response—demands a distributed computing model.
By balancing workloads across cloud and edge resources, the continuum enables low-latency processing close to data sources while ensuring centralized coordination and long-term storage in the cloud.
This dynamic allocation of computing and intelligence not only enhances responsiveness and system resilience but also supports more efficient use of infrastructure and energy—two key concerns in large-scale, heterogeneous urban environments.

In this paper, we present a real-world use case of intelligent climate monitoring inside an office building, implemented within a cloud–edge continuum and integrated with a smart city data space.
The use case demonstrates how edge computing, real-time analytics, and decentralized data sharing can be applied to optimize indoor environmental conditions, improve energy efficiency, and enhance occupant comfort.
By deploying AI/ML services close to the data source and enabling secure collaboration through data space principles, this implementation offers a practical and scalable model for building-level intelligent management.
It serves as a concrete example of how cloud-native and edge-native technologies, combined with data sovereignty frameworks, can contribute to the broader transition toward resilient, data-driven smart city ecosystems.


\section{The Paradigm of Sovereign Data Spaces}
The design of intelligent urban systems is undergoing a significant architectural transformation, driven by the convergence of sovereign data ecosystems, distributed computing infrastructures, and privacy-preserving artificial intelligence. This section surveys the state-of-the-art across these domains, establishing the technological foundations for harnessing data within complex smart city environments.

The contemporary digital economy is moving away from monolithic, centralized data platforms towards decentralized, sovereign data ecosystems \cite{curry2023data}.
Traditional data lakes, while powerful, have often created new data silos and concentrated market power, thereby hindering innovation \cite{otto2022federated}.
Data spaces have emerged as a federated alternative, enabling organizations to share and jointly utilize data while preserving the data sovereignty of participants.
This paradigm shift is not merely a technological evolution but a response to structural limitations of centralized platforms, aiming to establish a more equitable and collaborative data economy as envisioned by Europe's data strategy.
The operation of a data space is defined by a set of core principles enforced through technical, legal, and organizational measures. 
First, data sovereignty ensures that data owners retain ultimate and continuous control over their assets, technically enforced via usage control policies embedded within data transactions.
Second, multi-level interoperability (spanning technical, semantic, and legal layers) is essential to break down data silos, with standardized data models providing a \textit{lingua franca} for seamless data integration. 
Finally, trust is established ``by design'' through a unified framework of transparent governance, legally binding agreements, and certification of participants and software components.

\subsection{Foundational Blueprints and Technological Convergence}
The abstract concept of a data space is being operationalized by several key European initiatives that provide complementary, layered contributions to the ecosystem. 
The International Data Spaces Association (IDSA) offers one of the most mature blueprints, the IDS Reference Architecture Model (IDS-RAM), whose core technical component is the IDS Connector—a standardized software gateway that technically enforces data sovereignty at the participant level \cite{idsa_ram}.
In parallel, Gaia-X provides a broader, higher-level vision for a federated and transparent European data and cloud infrastructure, establishing a common framework of policies and rules to link independent cloud and edge providers into an interoperable ecosystem \cite{gaia-x_arch}. 
While Gaia-X provides governance and IDSA provides the secure endpoint, FIWARE delivers practical, open-source building blocks, such as the Context Broker and the ETSI-standardized NGSI-LD API, which are particularly well-suited for realizing data spaces in real-time, data-intensive domains like smart cities \cite{bauer2019ngsi}.
Recognizing the risk of fragmentation, these initiatives, along with the Big Data Value Association (BDVA), formed the Data Spaces Business Alliance (DSBA) to drive technical convergence and develop a common reference model, ensuring that their respective components and standards are compatible \cite{dsba_convergence}.

\subsection{Core Protocols for Interoperable Data Exchange}
The practical implementation of data spaces relies on core standardized protocols. 
The Dataspace Protocol (DSP), currently undergoing standardization at the Eclipse Foundation, is emerging as the international standard for governing core interactions \cite{dsp_spec}. 
It is a suite of specifications defining the processes for discovering data offerings (Catalog Protocol), electronically negotiating usage terms (Contract Negotiation Protocol), and managing the subsequent data transaction (Transfer Process Protocol). 
While the DSP standardizes the negotiation about data, the FIWARE NGSI-LD API provides a powerful, standardized interface for exchanging the content of real-time, dynamic data, which is the lifeblood of IoT environments.
Based on a property graph model, NGSI-LD uses a ``JSON-LD context'' to link human-readable attributes to globally unique URIs, enabling seamless semantic interoperability between heterogeneous systems \cite{bauer2019ngsi}. 
These protocols are complementary: the DSP governs permissions at a business and policy level, while NGSI-LD provides the technical mechanism for accessing the dynamic data content itself.

\subsection{The Cloud-Edge Continuum as a Sovereign-by-Design Infrastructure}
While data spaces provide the governance frameworks, the cloud-edge continuum provides the necessary distributed computational infrastructure for these systems to run effectively in demanding smart city contexts \cite{edgecloud_survey2025}. 
The sheer volume and velocity of IoT data, combined with the low-latency requirements of urban services, make a purely cloud-centric model untenable. 
The cloud-edge continuum enables a strategic balance, performing real-time processing at the network edge while leveraging the cloud for large-scale model training and long-term storage. 
This architectural choice is also a technical prerequisite for realizing data sovereignty in practice. 
Edge computing allows analytics and AI inference to occur at the source, on infrastructure trusted by the data owner. 
Consequently, participants can share the results of local processing (e.g., an anomaly alert) rather than the raw, sensitive source data.
This principle is extended further by Federated Learning (FL), a distributed machine learning technique where a global model is trained by aggregating updates from local models without the raw data ever leaving its trusted environment \cite{mcmahan2017communication, fl_survey2022}.
This approach enables collaborative intelligence while preserving data privacy by design, making it a key architectural pattern for achieving true data sovereignty.

\subsection{Applications in Intelligent Urban Environments}
These architectural and technological principles are being actively applied to intelligent infrastructure monitoring. 
The literature on smart cities is rich with examples of data-driven applications to enhance urban life, but a critical and persistent theme is the paramount importance of data governance, sovereignty, and security, particularly in multi-stakeholder environments \cite{haque2022conceptualizing, ali2022security}. 
Data space architectures are designed precisely to solve these challenges. 
Within this context, intelligent building management has become a key application area for AI. 
For instance, research has demonstrated the feasibility of using machine learning models to predict indoor $CO_2$ concentrations to optimize ventilation, thereby improving air quality while reducing energy consumption \cite{taheri2021learning}.
While prior work often focused narrowly on algorithmic aspects, a holistic approach requires integrating such intelligent services into a scalable, secure, and sovereign framework, addressing the broader data governance and interoperability challenges that are critical for real-world deployment.




\section{Smart Building Datasets}

We base our work on three distinct sets of data: 
\begin{itemize}
    \item The first dataset, includes $CO_2$ concentration,  temperature, and relative humidity data from 5 apartments (16 rooms), as well as the type of the room (e.g., kitchen, office, bedroom, etc.), collected over a period of 8 months in Denmark,  presented in \cite{noauthor_dataset_nodate}~\&\cite{andersen_exploring_2024}.
    \item The second dataset is available on kaggle with measurements correlating environmental and $CO_2$ measurements with room occupancy~\cite{pooriamst2017occupancy}.
    \item The third dataset was collected from a smart office environment deployed in our own office space, designed to support intelligent indoor climate monitoring and control. It captures environmental conditions and occupancy indicators from multiple sensor nodes deployed across office rooms and shared spaces. Each sensor node includes multiple sensing modalities and records data at regular intervals (e.g., every 10 seconds to 1 minute, depending on sensor type). The core data streams include: (1) ambient air temperature measured at multiple points in the room, (2) relative humidity levels to assess thermal comfort and ventilation needs, (3) $CO_2$ concentration used as an indicator of air quality and occupancy intensity, (4) light intensity for detecting presence/activity and correlating with comfort, and (5) motion/occupancy captured via passive infrared (PIR) sensors. For each sensing point, we also include information for: (1) the unique room identifier of the physical space from which data is collected and (2) the device identifier for traceability and troubleshooting.
    
\end{itemize} 


\section{ML Models}

\subsection{Model Selection}
The initial phase of our work focused on selecting the most appropriate models to execute the 3 goals of our application: cleanup IoT sensor data, forecast future measurements, and finally detect the presence of people in the building in the upcoming period.
Our choice was to use an \texttt{IsolationForest} model for \textsf{anomaly detection}, a \texttt{sequential neural network} for \textsf{forecasting}, and a \texttt{RandomForestClassifier} for \textsf{presence detection}. 

\subsubsection{Anomaly Detection}
For the anomaly detection task, our goal was to identify unusual and potentially erroneous readings from the environmental sensors.
These readings include error values output by the sensor module (e.g., negative numbers that indicate sensor error) and values well outside the expected conditions of an indoor environment (e.g., a temperature value of 200 degrees centigrade).
We chose to use an \texttt{IsolationForest} model due to its efficiency and ability to isolate outliers in noisy, real-world sensor data.
Specifically, we build specialized models for each key environmental metric: temperature, humidity, and $CO_2$, as well as a combined model for all 3 types of data. 

\subsubsection{Forecasting}
To predict future environmental conditions, we developed a forecasting model based on a \texttt{sequential neural network}.
It was employed due to its power to learn the intricate temporal dependencies within time series data and make accurate predictions.
The core of this approach involved transforming the time-series data into sequences, where a set of past observations (specifically the last 3 consecutive readings) was used to predict the next value.
We structured our model with multiple dense layers and ReLU activation functions, designed to learn the temporal patterns within the data. 
The model was compiled using the \texttt{Adam} optimizer and the \texttt{mean squared error} loss function. 

\subsubsection{Presence Detection}
For presence detection, we selected a model based on the \texttt{RandomForestClassifier} to predict room occupancy based on the aforementioned environmental factors.
It was selected for its robustness and ability to effectively capture the complex, non-linear relationships between environmental variables and occupancy.

\subsection{Model Training \& Validation}

The training phase of our work focused on preparing the environmental time-series data and ML model training. 
This involved essential preprocessing steps, including merging datasets, cleaning, and normalization to ensure data quality.

\subsubsection{Anomaly Detection}
For the anomaly detection task, we used \texttt{IsolationForest} to build specialized models for each key environmental metric: temperature, humidity, and $CO_2$.
To optimize the performance of each model, we implemented a grid search to systematically test the model's hyperparameters and identify the most effective configuration.
The grid search explored a range of values for key hyperparameters, including the number of estimators ($n\_{estimators}$: [50, 100, 200]), the proportion of samples to draw ($max\_{samples}$: [0.6, 0.8, 1.0]), and the expected contamination rate ($contamination$: [0.01, 0.05, 0.1]) 
This process allowed us to fine-tune separate models for our application, achieving an accuracy of over 92\% during training.  
The best-performing estimator for each of these models was then serialized and deployed.

\begin{table}[h]
    \centering
     \caption{Performance metrics of the anomaly detection models.}
    \label{tab:test_anomaly_detection}
    \begin{tabular}{lcccc}
        \hline
        \textbf{Model} & \textbf{Accuracy} & \textbf{Precision} & \textbf{Recall} & \textbf{F1-score} \\
        \hline
        Combined & 0.90 & 0.98 & 0.90 & 0.94 \\
        Temperature & 0.90 & 0.98 & 0.89 & 0.94 \\
        Humidity & 0.91 & 0.98 & 0.90 & 0.95 \\
        Air Quality & 0.90 & 0.97 & 0.89 & 0.94 \\
        $CO_2$ & 0.91 & 0.96 & 0.90 & 0.94 \\
        \hline
    \end{tabular}
\end{table}

\subsubsection{Forecasting}
After splitting the data into training and testing sets, the model was trained for a total of 50 epochs, and to ensure the selection of the most effective model, we implemented a checkpoint system. This approach saved the model that achieved the best Mean Absolute Error (MAE) for each feature at any point during the training, such as selecting a model from the 30th epoch if it outperformed later ones. We chose MAE as the evaluation metric because it is inherently less sensitive to outliers, providing a more robust and reliable measure of model performance for our specific dataset. To further fine-tune the model, we conducted hyperparameter tuning, focusing on the $learning~rate$ with values such as [0.01, 0.001, 0.0001] and the number of $neurons$ in each dense layer, exploring configurations like [ 64, 128, 256].
This process led to very low \textsf{Mean Absolute Error} (MAE) values across all models: less than 0.01 for temperature, less than 0.03 for humidity, less than 15 for $CO\_2$, and less than 3 for the combined model.
Its performance was validated on the test set, and the final trained model was deployed, enabling us to generate future predictions for sensor readings based on recent historical data.

\begin{table}[h]
    \centering
     \caption{Performance metrics of sequential models.}
    \label{tab:test_forecasting}
    \begin{tabular}{lcccc}
        \hline
        \textbf{Model} & \textbf{MAE} & \textbf{RMSE} & \textbf{MAPE}\\
        \hline
        Combined & 3.1683 & 10.0553 & 0.1264  \\
        Temperature & 0.0115 & 0.1426 & 0.0005  \\
        Humidity & 0.0328 &0.1414 &  0.0008  \\
        Air Quality & 0.5884 & 2.8617 & 0.0067  \\
        $CO_2$ & 16.2783 & 32.1588 & 0.0237  \\
        \hline
    \end{tabular}
\end{table}

\subsubsection{Presence Detection}
For this part of the application, we merged the first two datasets we used to create a comprehensive training set.
The combined temperature, humidity, $CO_2$, and occupancy data were split into training and testing sets, and the model was trained on the former.
We used a \texttt{RandomForestClassifier} and to optimize it, we implemented a grid search to tune its hyperparameters. This process involved testing configurations for parameters such as the number of trees in the forest ($n\_estimators$), the maximum depth of each tree ($max\_depth$), and the minimum number of samples required to split a node ($min\_samples\_split$).
Each model achieved training accuracy over 95\%

\begin{table}[h]
    \centering
     \caption{Performance metrics of the Presence Detection Model.}
    \label{tab:test_presence}
    \begin{tabular}{lcccc}
        \hline
        \textbf{Model} & \textbf{Accuracy} & \textbf{Precision} & \textbf{Recall} & \textbf{F1-score} \\
        \hline
        Random Forest Classifier & 0.94 & 0.95 & 0.97 & 0.96 \\
        \hline
    \end{tabular}
\end{table}

\subsection{Presence Detection Evaluation}

Table \ref{tab:eval_presence} presents the performance metrics of the Presence Detection model evaluated on the real-time data collected from the Smart Office environment for the two classes defined (0:Unoccupied and 1:Occupied).
For the Unoccupied class, the model achieved an accuracy of 72\%, with a precision of 0.77, a recall of 0.83, and an F1-score of 0.80, based on 10032 samples. For the Occupied class, the model's performance was notably lower, with an accuracy of 71.9\%, a precision of 0.59, a recall of 0.5, and an F1-score of 0.54, using 4968 samples. 
The weighted average across both classes indicates an overall accuracy of 72\%, with a precision of 0.71, recall of 0.72, and F1-score of 0.71, computed over a total of 15000 samples.

\begin{table}[h]
    \centering
     \caption{Performance metrics of the Presence Detection Model on the Smart Office real-time data.}
    \label{tab:eval_presence}
    \begin{tabular}{lccccc}
        \hline
        \textbf{Label} & \textbf{Accuracy} & \textbf{Precision} & \textbf{Recall} & \textbf{F1-score} & \textbf{Samples} \\
        \hline
        Unoccupied & 0.720 & 0.77 & 0.83 & 0.80 & 10032 \\
        Occupied  & 0.719 & 0.59 & 0.50 & 0.54 & 4968 \\
        Weighted avg & 0.72 & 0.71 & 0.72 & 0.71 & 15000 \\
        \hline
    \end{tabular}
\end{table}

These results indicate that the classifier model trained is effective in detecting unoccupied states, as shown by the high recall and F1-score for the Unoccupied class. 
This suggests strong performance in correctly identifying times when no presence is detected, which is beneficial for energy-saving and occupancy-monitoring applications.
However, it model struggles to accurately detect the Occupied state, with a substantially lower recall and F1-score.
This implies a higher rate of false negatives—failing to detect presence when it actually exists. 
This may limit our solution's reliability in scenarios requiring real-time awareness of human activity.
The performance gap between the two classes points to potential improvements, such as rebalancing on the training data, tuning model hyperparameters, or incorporating additional contextual features to enhance the classifier's ability to identify occupied conditions more accurately.
Also, it may be due to differences in the sensor characteristics in use in the 2 datasets we used for training and the one we use in our installation.
Partial retraining of our base model on the data collected by our own data source could be a solution to increase the scores achieved, while not overfitting it to our dataset.

\section{Smart Building Application}
\begin{figure}[h]
    \centering
    \includegraphics[width=0.95\linewidth]{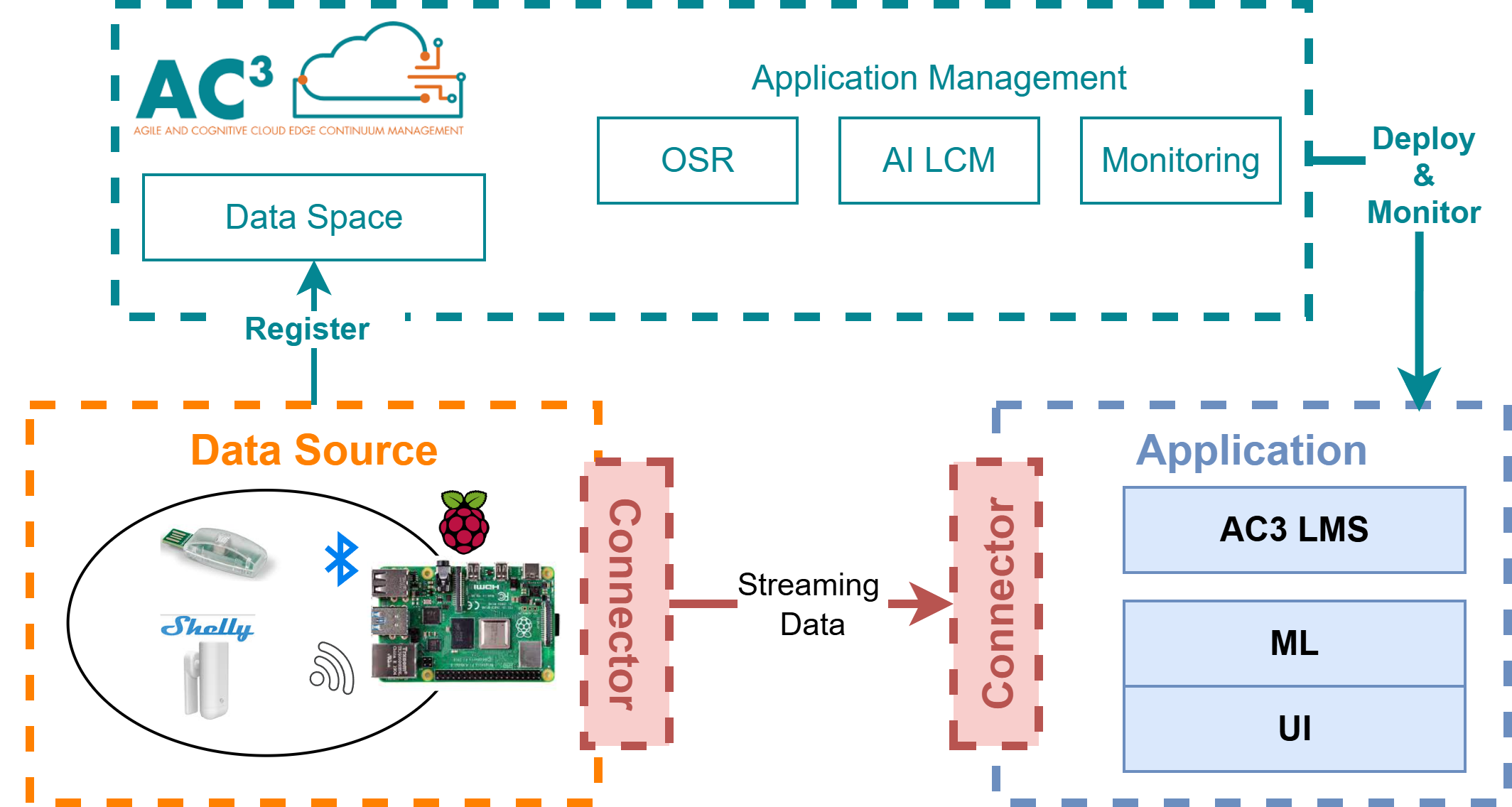}
    \caption{Schematic representation of the smart building application's deployment using $AC^3$ and our deployed data source.}
    \label{fig:depl-arch}
\end{figure}

Figure~\ref{fig:depl-arch} presents a schematic representation of the deployment architecture for our smart building application using the \accc framework. 
The diagram illustrates how data collected from our smart office sensor deployment is registered as a data source into a  secure, interoperable Data Space provided by the \accc project.
This data source is then used by our application, deployed and monitored by the \accc project in our edge or cloud testbeds.


\subsection{Data Source}
Our IoT deployment is registered as a Data Source to the \accc Data Space.
This registration involves creating a semantic representation of the data source using standardized GAIA-X vocabularies to describe its characteristics, such as device type, location, data modality, and communication protocols. 
Each data source is annotated with metadata including a unique \texttt{Asset ID}, which serves as its persistent identifier within the Data Space, and an \texttt{endpoint address} that defines where the data can be accessed from.
This information is structured using linked data principles and published to the Data Space catalogue, enabling data consumers to discover, query, and retrieve streaming data in an interoperable and secure manner. 
The metadata also includes temporal resolution, update frequency, licensing terms, and access policies, ensuring both transparency and compliance with data governance frameworks such as GAIA-X and IDSA.

To access the registered data, applications must interact with the Data Space through a data connector based on the Eclipse Data Connector (EDC) framework. 
These connectors are deployed at both the data provider and data consumer ends, establishing a trusted, policy-enforced communication channel. 
The connectors handle the negotiation of data contracts, enforcement of access control rules, and secure data transfer, ensuring compliance with the defined usage conditions.
This mechanism decouples the data source from consuming applications, allowing secure and scalable integration of third-party services and applications into the smart building ecosystem.

\subsection{Cloud-Edge Deployment}
The deployment of the smart building application is managed through the \accc platform, which provides a flexible and automated mechanism for orchestrating applications across cloud and edge environments.
\accc supports the full application lifecycle—from registration and configuration to deployment, monitoring, and updates—through its AI Life-Cycle Management (LCM).
Applications are packaged as composable services and described using deployment descriptors that define their resource requirements, placement preferences, and data dependencies.
Based on these descriptors and contextual information (e.g., network conditions, latency requirements, or resource availability), \accc dynamically deploys the application components either at the edge—close to the data sources for low-latency processing—or in the cloud to leverage greater compute capacity.
The platform ensures that all deployed services are properly connected to the necessary data streams and continuously monitors their performance and health, enabling adaptive scaling, fault recovery, and efficient use of infrastructure resources.

\subsection{Interface}
The Smart Building Monitoring Application is a web-based system developed using the Python Flask framework, designed to provide real-time and historical insights into occupancy and environmental conditions within a building.
It serves as a user-friendly interface for facility managers, researchers, and end-users who need to monitor and understand how various indoor environments are used and how their conditions change over time.

\begin{figure}[h]
    \centering
    \includegraphics[width=0.95\linewidth]{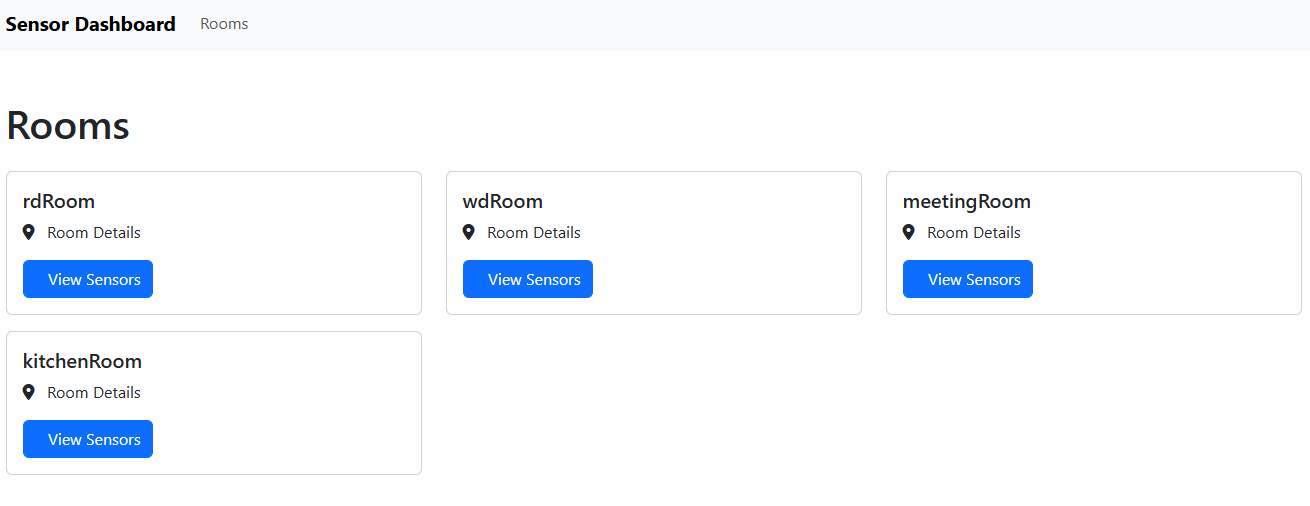}
    \caption{The smart building application's room list page.}
    \label{fig:view-rooms}
\end{figure}

Upon logging into the application, users are presented with a dashboard (Figure~\ref{fig:view-rooms}) that lists all the monitored rooms in the building (e.g., \texttt{rdRoom}, \texttt{wdRoom}, \texttt{meetingRoom}, \texttt{kitchenRoom}).
Each room tile includes key metadata and a button to "View Sensors", allowing users to drill down into room-specific information.

\begin{figure}[h]
    \centering
    \includegraphics[width=0.95\linewidth]{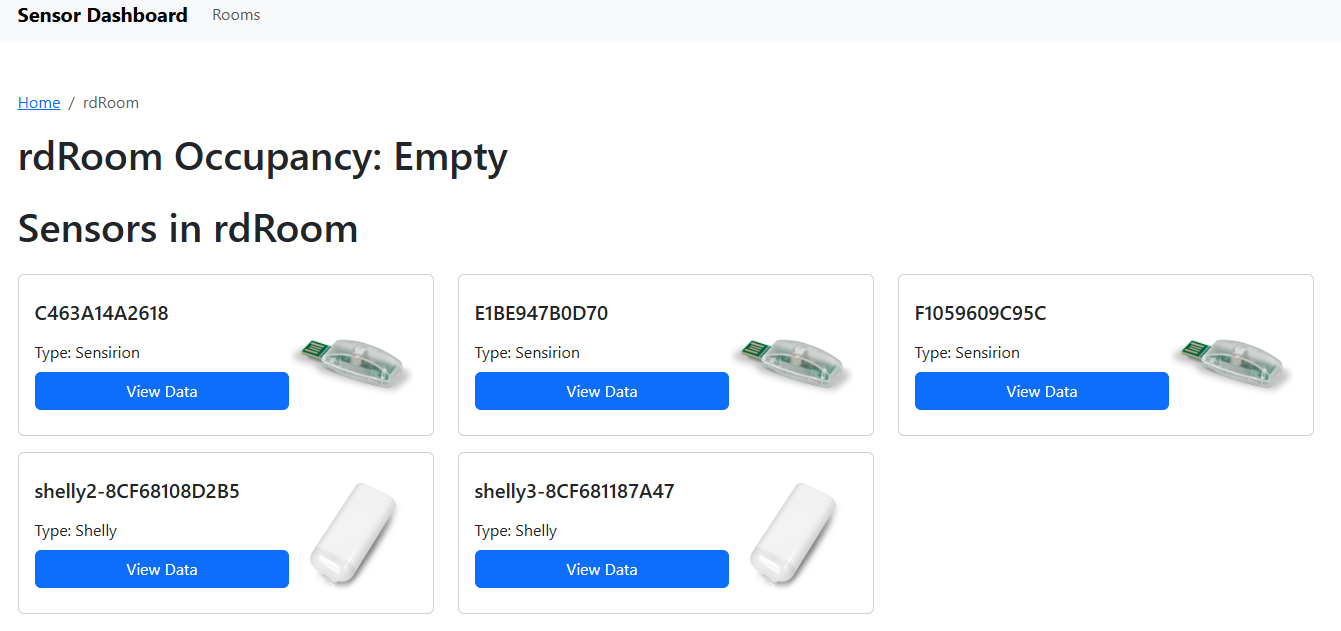}
    \caption{The smart building application's room information page.}
    \label{fig:view-room}
\end{figure}

\begin{figure}[h]
    \centering
    \includegraphics[width=0.95\linewidth]{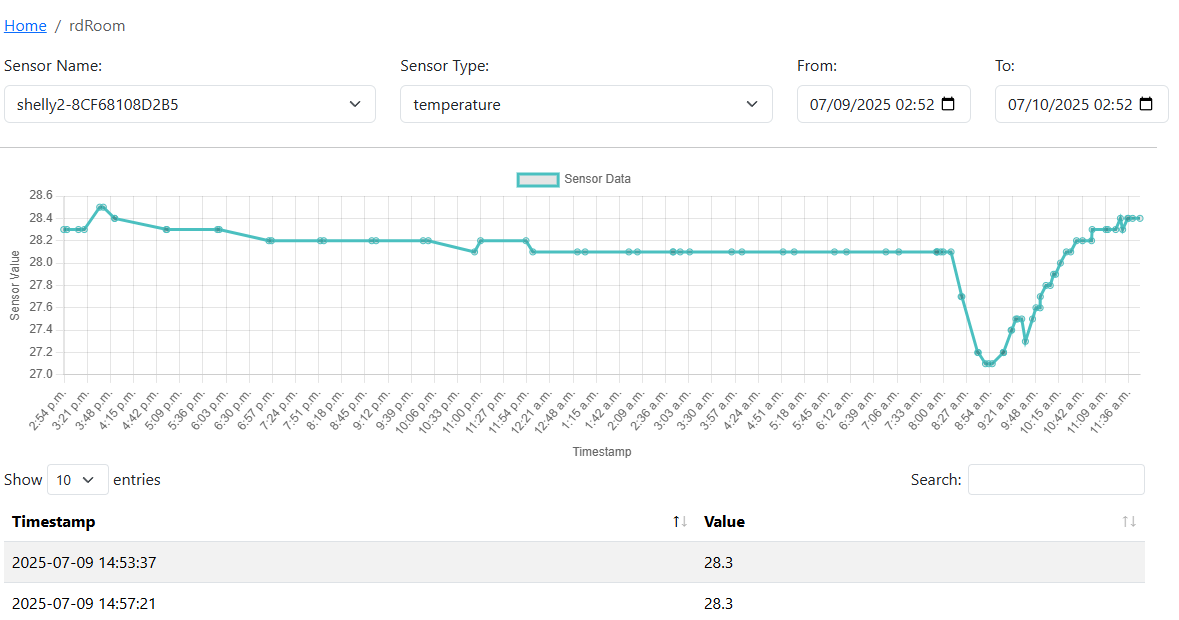}
    \caption{The smart building application's sensor data view page.}
    \label{fig:view-data}
\end{figure}

Each room view (Figure~\ref{fig:view-room}) displays the current occupancy status (e.g., \textsf{Occupied} or \textsf{Empty}) and lists all associated sensors. 
These sensors can be of different types, such as \textsf{Sensirion} (used for environmental measurements like temperature and humidity) or \textsf{Shelly} (often used for motion or door sensing). 
Each sensor is identified by a unique ID (e.g., \texttt{shelly2-XYZ}) and a button that allows users to view detailed sensor data.
When a specific sensor is selected, users are taken to a detailed analytics view (Figure~\ref{fig:view-data}) that displays:
\begin{itemize}
    \item The sensor name and type of measurement (e.g., temperature).
    \item A selectable time window for querying data.
    \item A dynamic line chart plotting sensor values (e.g., temperature) over time.
    \item A data table showing raw values and timestamps for each measurement.
\end{itemize}

This combination of graphical and tabular data allows users to visually detect trends, anomalies, or changes in environmental conditions—such as a drop in temperature during unoccupied periods or a spike during peak usage.

\section{Conclusion}

This paper demonstrated how data spaces, in combination with edge-enabled IoT and AI services, can be effectively harnessed to build intelligent and resilient smart building infrastructures. 
Using the \accc platform, we deployed a complete pipeline that integrates semantically described sensor data, sovereign data exchange via EDC-based connectors, and cloud–edge orchestration of modular AI applications for real-time environmental monitoring and decision-making.

Our smart building use case highlights several key advantages of this architecture: secure and interoperable data integration using Data Space principles, flexible application deployment across the cloud–edge continuum, and the seamless incorporation of AI/ML models for forecasting, anomaly detection, and occupancy prediction. 
The \accc framework enabled dynamic, context-aware deployment and lifecycle management of these services, while ensuring continuous monitoring and adaptability to changing environmental and operational conditions.

By aligning with emerging European standards such as GAIA-X and IDSA, our approach ensures data sovereignty, reusability, and compliance—critical requirements for scalable smart city systems. This work not only validates the technical feasibility of federated, AI-powered smart building applications but also provides a blueprint for replicable deployments in broader urban contexts.

Looking forward, future work will focus on integrating additional urban verticals (e.g., energy optimization, mobility, and citizen engagement), supporting federated learning across distributed buildings, and further automating the orchestration of complex data-driven services across heterogeneous infrastructures. 
The continued evolution of data space standards and edge intelligence will be essential to unlocking the full potential of data-centric smart cities.

\bibliographystyle{IEEEtran}
\bibliography{conference_101719}

\end{document}